# Daily detection and quantification of methane leaks using Sentinel-3: a tiered satellite observation approach with Sentinel-2 and Sentinel-5p

Sudhanshu Pandey[1,2], Maarten van Nistelrooij[1], Joannes D. Maasakkers[1], Pratik Sutar[1], Sander Houweling[3], Daniel J. Varon[4], Paul Tol[1], David Gains[5], John Worden[2], Ilse Aben[1,3]

[1] SRON Netherlands Institute for Space Research, Leiden, the Netherlands
[2] Jet Propulsion Laboratory, California Institute of Technology, Pasadena, CA, USA
[3] Department of Earth Sciences, Vrije Universiteit Amsterdam, Amsterdam, the Netherlands
[4] School of Engineering and Applied Sciences, Harvard University, Cambridge, United States
[5] GHGSat, Inc., Montréal, H2W 1Y5, Canada

*Correspondence to*: Sudhanshu Pandey (sudhanshu.pandey@jpl.nasa.gov)

**Abstract.** The twin Sentinel-3 satellites have multi-band radiometers which observe at methane-sensitive shortwave infrared bands with daily global coverage and 500 m ground pixel resolution. We investigate the methane observation capability of Sentinel-3 and how its coverage-resolution combination fits between Sentinel-5p and Sentinel-2 within a tiered observation approach for methane leak monitoring. Sentinel-5p measures methane with high precision and daily global coverage allowing worldwide leak detection, but it has a coarse spatial resolution of 7 km × 5.5 km. The Sentinel-2 twin satellites have multi-band instruments that can identify source locations of major leaks (> 1 t/h) with their methane observations of 20 m resolution under favorable observational conditions, but these satellites lack daily global coverage.

We show that methane plume enhancements can be retrieved from the shortwave infrared bands of Sentinel-3. We report lowest emission detections by Sentinel-3 in the range of 8-20 t/h, depending on location and wind conditions. We demonstrate Sentinel-3's identification and monitoring of methane leaks using two case studies. Near Moscow, Sentinel-3 shows that two major short-term leaks, separated by 30 km, occurred simultaneously at a gas pipeline and appear as a single methane plume in Sentinel-5p data. For a major Sentinel-5p leak detection near the Hassi Messaoud oil/gas field in Algeria, Sentinel-3 identifies the leaking facility emitting continuously for 6 days, and Sentinel-2 pinpoints the source of the leak at an oil/gas well. Sentinel-2 and Sentinel-3 also show the 6-day leak was followed by a four-month period of burning of the leaking gas, suggesting a gas well blowout to be the cause of the leak. We find similar source rate quantifications from plume detections by Sentinel-3 and Sentinel-2 for these leaks, demonstrating Sentinel-3's utility for emission quantification. These case studies show that zooming in with Sentinel-3 and Sentinel-2 in synergy allows precise identification and quantification as well as monitoring of the sources corresponding to methane anomalies observed in Sentinel-5p's global scans.

# 1 Introduction

Methane ($CH_4$) is the second most important anthropogenic greenhouse gas after carbon dioxide. Its atmospheric abundance has increased more than 250 % since preindustrial times. Due to its strong global warming potential, it is responsible for at least 25 % of anthropogenic radiative forcing (IPCC, 2021). More than half of the world's methane emissions are from anthropogenic sources like oil and gas facilities, coal mines, waste management, domestic ruminants, and agriculture (Saunois et al., 2020). Reducing these emissions has been recognized as an effective opportunity for climate change mitigation in the short term: Reduction by as much as 45 per cent is possible within a decade using cost-effective, existing technologies, which would avoid nearly 0.3°C of global warming by 2045 and would be a strong contribution to the Paris Climate Agreement's goal to limit global temperature rise to 1.5˚C (Nisbet et al., 2020; CCAC, 2022; UNEP, 2021).

Super-emitter events are large leaks from methane point sources caused by accidents, malfunctioning equipment, or abnormal operating conditions. Many studies have shown that a few super-emitters are often responsible for a disproportionately large fraction of total regional emissions (Brandt et al., 2016; Frankenberg et al., 2016; Zavala-Araiza et al., 2017; Duren et al., 2019; Cusworth et al., 2021; Cusworth et al. 2022). Fixing methane super-emitter leaks is a cost-effective and practical means of climate change mitigation. Here, we investigate the methane leak observation capability of the multi-band Sentinel-3 satellites, which observe at 500 m resolution with daily global coverage. We also assess the additional benefit of using Sentinel-3 with Sentinel-5p and Sentinel-2 observations in a tiered observation approach.

The TROPOspheric Monitoring Instrument (TROPOMI) aboard the ESA's Sentinel-5p satellite measures column-averaged methane with high precision and daily global coverage (Lorente et al., 2022; Veefkind et al., 2012). Sentinel-5p regularly detects large methane plumes across the globe (Pandey et al., 2019; Varon et al., 2019; Lauvaux et al., 2021; Sadavarte et al., 2021; Maasakkers et al., 2022a; Maasakkers et al., 2022b; Cusworth et al., 2021), but it generally cannot pinpoint the sources of these plumes because of its relatively coarse spatial resolution of 5.5 km × 7 km. Maasakkers et al. (2022b) addressed this problem by using a plume-rotation technique on multiple overpasses of Sentinel-5p data to get a rough source location area (within a few km) and then zoom-in with the high-resolution GHGSat satellites (25–50 m). This plume-rotation technique can only be applied to persistent sources for which there are multiple plume observations with varying wind directions. The GHGSat satellites are specifically designed for methane point-source detection, but these satellites have a limited observation area of 12 km × 12 km (Jervis et al., 2021) and only a small fraction of the data is publicly available (for example, through ESA's third-party mission program). Hyperspectral satellite instruments with 5-10 nm spectral resolution and high spatial resolution (~30 m), like China's ZY1 and Italian Space Agency's PRISMA, are able to detect methane plumes from super-emitters near the source under favorable observational conditions (Irakulis-Loitxate et al., 2021, Guanter et al., 2021, Cusworth et al., 2021). However, the revisit times of these satellites are limited by spatial coverage and tasking requirements (Jacob et al., 2022).

Multi-band satellites such as ESA's Sentinel-2 and NASA's Landsat have similar spatial resolution to the hyperspectral and GHGSat satellites, but they have rapid revisit times: global coverage in < 5 days for Sentinel-2, and < 3 days for the combination of Sentinel-2, Landsat-8, and Landsat-9 (Li & Chen, 2020). Despite the very low spectral resolution of these instruments in short wave infrared (SWIR) bands (100-200 nm), it is nevertheless possible to retrieve methane plumes in close vicinity of massive point sources using a combination of their SWIR observations at 1600 nm and 2200 nm, referred to respectively as SWIR-1 and SWIR-2 bands from here on (Varon et al., 2021). Ehret et al. (2022) used Sentinel-2 for methane leak monitoring of oil/gas facilities at 7000 geographical sites in four countries. Irakulis-Loitxate et al. (2022) performed a historical assessment of 29 super-emitters at oil/gas facilities in West Turkmenistan using Sentinel-2, PRISMA, and Landsat guided by Sentinel-5p.

Sentinel-3 is a European Earth observation satellite mission developed for ocean, land, atmospheric, emergency, security, and cryospheric applications (Donlon et al., 2012). The Sentinel-3 mission consists of two satellites, Sentinel-3A and Sentinel-3B, that measure in the SWIR bands at 500 m × 500 m resolution with a swath width of 1420 km, enabling daily global coverage by combining the two. Although Sentinel-3 is not designed to measure methane, its SWIR bands are sensitive to absorption by methane molecules in the atmosphere.

Here, we investigate the methane plume detection and source rate quantification capabilities of Sentinel-3. We assess the impact of methane column enhancements on Sentinel-3 SWIR observations and use the multi-band multi-pass method (Varon et al., 2021) to retrieve methane enhancements (Section 2). To demonstrate the capability of Sentinel-3 to detect large leaks, we present two case studies in Section 3: a gas pipeline in Russia and an oil/gas facility in Hassi Messaoud, Algeria. We use Sentinel-3 and Sentinel-2 to identify the facilities responsible for the large methane plumes detected by Sentinel-5p at these locations and provide additional observational constraints. We then discuss the emission detection limit of Sentinel-3 (Section 4). In Section 5, we explore the role of the satellite in a tiered observation approach for monitoring methane leaks.

## 2 Methane observation with Sentinel-3

Here we describe the Sentinel-3 mission and the methane retrieval method. The Sentinel-5p and Sentinel-2 satellites are described in the supplementary information (SI) Section 1.

### 2.1 Sentinel-3 satellites

The main objective of the Sentinel-3 mission is to measure sea surface topography, sea and land surface temperature, and ocean and land surface color to support ocean forecasting systems, and environmental and climate monitoring (Donlon et al., 2012). Like Sentinel-2, Sentinel-3 is a twin satellite mission consisting of Sentinel-3A (launched on 16 February 2016) and Sentinel-3B (launched on 25 April 2018) flying 140 degrees out of phase of each other. Sentinel-3 satellites are designed for a 7-year operational lifetime with sufficient fuel for up to 12 years of continuous operations. Each satellite has an identical multiple instrument payload including a push-broom imaging spectrometer called the Ocean and Land Color Instrument (OLCI) and a dual view (near-nadir and inclined) conical imaging radiometer called the Sea and Land Surface Temperature Radiometer (SLSTR) instrument. The SLSTR has two methane-sensitive SWIR bands, near methane's 1.65 and 2.3 µm absorption features (Figure 1). The principal objective of SLSTR products is to provide global and regional sea and land surface temperature. SLSTR observations have a spatial resolution of 500 m in the visible and SWIR channels and a swath width of 1420 km in nadir mode, which enables daily global coverage by combining both Sentinel-3 satellites. We refer to the Sentinel-3 SLSTR instrument as just "Sentinel-3" from here on.

## 2.2 Methane retrieval from Sentinel-3

We use Sentinel-3 SLSTR level-1B near real time top-of-atmosphere (TOA) radiances in the SWIR bands to retrieve methane column enhancements. These data are publicly available on the European Space Agency's Copernicus Open Access Hub as ortho-geolocated surface tiles on a regular quasi-Cartesian grid.

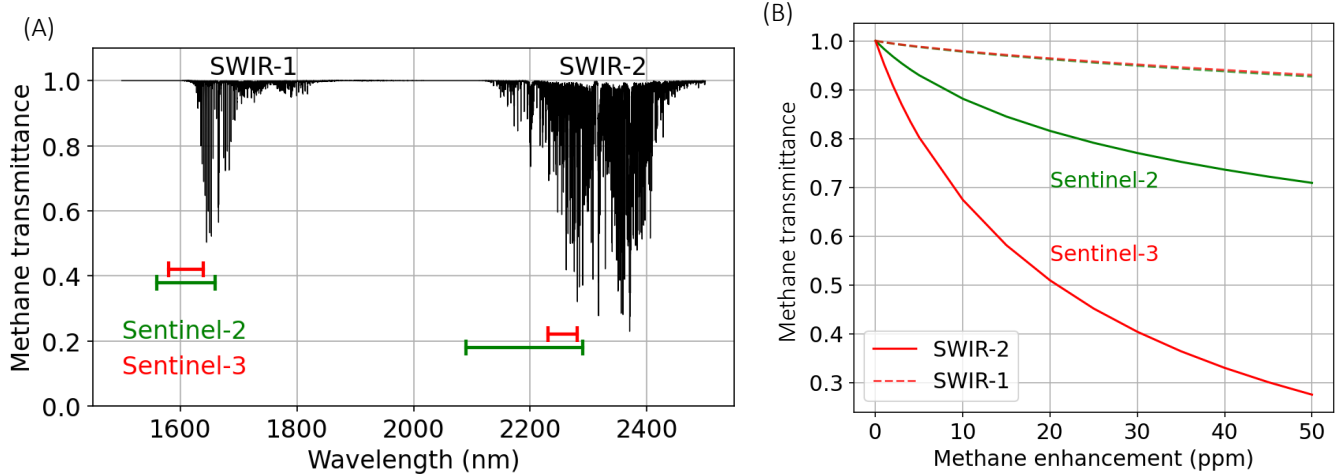

**Figure 1.** (A) Sentinel-3 and Sentinel-2 SWIR band widths, and the transmittance for 1 ppm methane enhancement in the 1500–2500 nm spectral range at a resolution of 0.025 nm. These satellites observe only the integrated signal per band. (B) Methane transmittance integrated over the SWIR-2 and SWIR-1 bands of Sentinel-3 and Sentinel-2 as a function of pixel methane enhancement.

Figure 1.A shows the transmittance for 1 ppm of column methane enhancement along with the SWIR band widths of Sentinel-3 and Sentinel-2. The SWIR bands of Sentinel-3, 1580–1640 nm (SWIR-1) and 2230–2280 nm (SWIR-2), lie within the methane absorption regions. Varon et al. (2021) showed that methane concentration enhancements in a plume can be retrieved from multi-band satellites like Sentinel-2 with the multi-band multi-pass (MBMP) method. We use the same method to retrieve methane enhancements from Sentinel-3 and Sentinel-2 but with small modifications, which are described in SI Section 2. The MBMP method calculates a fractional methane absorption signal image using the SWIR-2 TOA radiance measurements as the primary methane signal. Surface albedo artifacts in the SWIR-2 image are corrected using a set of reference SWIR images including the SWIR-1 band

image from the observation day and plume-free SWIR-1 and SWIR-2 images from another single overpass (when a plume-free overpass is available) or a combination of overpasses.

To retrieve a methane plume for a site, we consider a full month of Sentinel-3 observations around the observation day for an area of 1° × 1° around the site. We prepare the reference SWIR images by taking the pixel-wise median of five different overpasses. These overpasses are selected based on the five highest correlations in the 554 nm visible and near-infrared (VNIR) band images with the observation day with a correlation threshold of 0.70. This band is not influenced by the presence of methane plumes, hence it is useful to identify overpasses with surface features similar to that of the observation day. Selecting the reference images from overpasses near the observation days avoids MBMP artifacts due to seasonal changes in surface features such as vegetation growth. The sampling grids of Sentinel-3 data are spatially shifted between different daily overpasses, so we resample Sentinel-3 data to a common 0.001° × 0.001° (~ 100 m × 100 m) grid, which allows comparison of different overpass images. We find oblique stripes in the MBMP images of some scenes. These stripes are not aligned with the vertical or horizontal direction of the image data array. To remove oblique stripes, we determine the stripe angle using a two-dimensional fast Fourier transform and align stripes with the horizontal direction by image rotation. Row-wise medians are then subtracted from the rotated image. We also implement a feature-masking method based on local values of structural similarly index (Wang et al., 2004) to remove noise due to surface features and clouds observed in the methane-free VNIR bands. We mask out water pixels from our scenes using the water flag from the Sentinel-3 product. The destriping and feature-masking procedures are described in more detail in SI section 4.

The MBMP fractional absorptions are converted to methane enhancements (ppm) using a 25-layer, clear-sky radiative transfer model, which simulates radiances in the SWIR bands of Sentinel-3 and Sentinel-2 (see SI Section 2). We identify a plume in a methane enhancement image by first applying a median filter for spatial smoothing followed by a 95-percentile mask. We perform visual checks against Sentinel-3 VNIR images or visual surface imagery (e.g., Google Earth) to ensure that surface features or smoke plumes are not misinterpreted as methane plumes.

To calculate source rates for detected Sentinel-3, Sentinel-2, and Sentinel-5p plumes, we use the cross-sectional flux (CSF) method, described in SI Section 3, with fifth generation ECMWF atmospheric reanalysis (ERA5; Hersbach et al., 2020) wind data. Because low wind speeds favor plume detection, many of the plumes presented in this study are observed in such conditions. However, the source rate quantifications for low-wind plumes are difficult because low winds often cause rotating or blob-like plumes. Additionally, wind speed and direction estimates from reanalysis datasets like ERA-5 are very uncertain for low-wind conditions. We estimate source rates for plumes using the CSF method only when the wind speeds are higher than 2 m/s, which is also recommended by Varon et al. (2018).

Figure 2 shows an example of a methane plume retrieval from Sentinel-3 data. The corresponding unmasked methane concentration enhancement image for the scene is shown in SI Figure 1. The plume is observed on June 29, 2018 and originates from near the Korpezhe compressor station in Turkmenistan. The Sentinel-3 plume extends more than 10 km northwards from the leak site. Sentinel-2 observes the same leak 45 mins later. Varon et al. (2019) reported first methane plume detections from this site using GHGSat, and Varon et al. (2021) investigated the plume record of this site from Sentinel-2. The CSF-based source rate estimates for the plumes agree within the uncertainties: 76 ± 32 t/h (mean ± 1 standard deviation error) for Sentinel-3 and 56 ± 24 t/h for Sentinel-2.

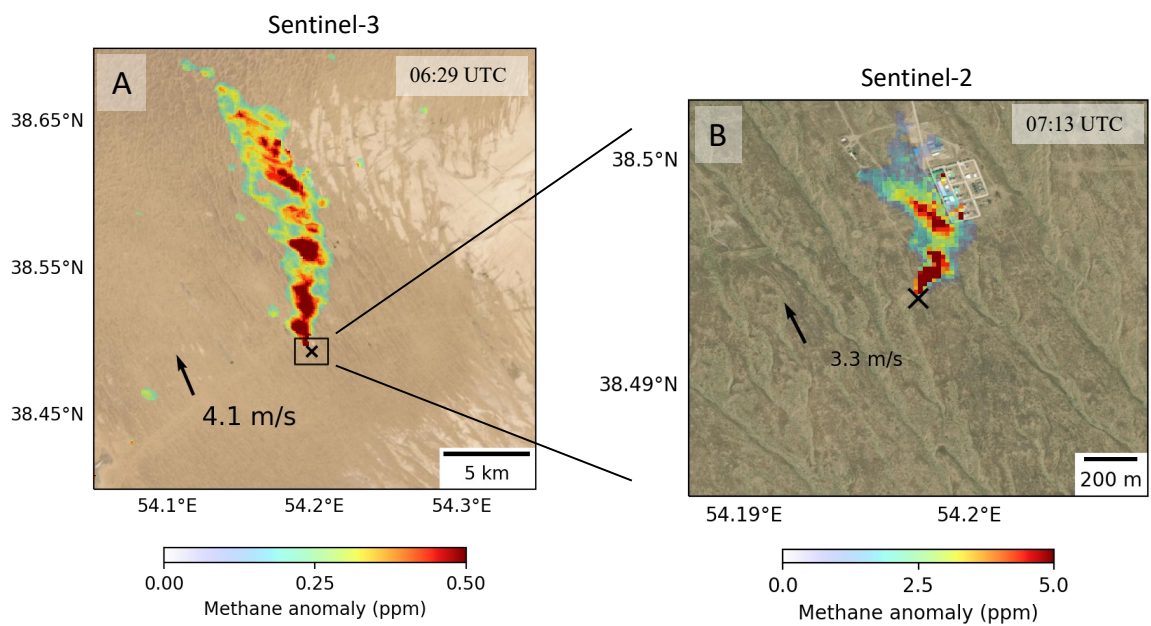

**Figure 2.** Observation of a methane plume by Sentinel-3 (A) and Sentinel-2 (B) on 29 June, 2018 from near the Korpezhe compressor station point source (black 'x'; location: 38.4939° N, 54.1977° E) in Turkmenistan. The area of Panel B is marked by the black box in Panel A. The plumes are overlaid on ESRI surface imagery.

# 3 Case studies

Massive methane plumes due to leaks in oil/gas infrastructure and landfills are regularly detected in Sentinel-5p global scans (e.g., Lauvaux et al., 2021; Maasakkers et al., 2022b). Here, we demonstrate a tiered satellite observation approach for methane leaks by using Sentinel-3 and Sentinel-2 to identify and monitor the plume sources of two Sentinel-5p plume detections.

## 3.1 Pipeline emissions in Russia

Figure 3 shows a large methane plume detected by Sentinel-5p at 10:20 UTC, June 18, 2021, over Russia, about 100 km northeast of Moscow. The highest column methane enhancement in the plume is ~500 ppb. The plume is oriented northeast, consistent with the wind direction. Sentinel-3 observes the scene at 08:42 UTC, 1 hour and 40 minutes before Sentinel-5p, and reveals that not one but two distinct large leaks, 30 km apart, fall within the Sentinel-5p plume. The locations of the two leaks are aligned with the wind direction, resulting in what appears as a single plume in Sentinel-5p observations. We zoom-in on the individual Sentinel-3 source locations with Sentinel-2, which observed the scene 12 minutes after Sentinel-3. The Sentinel-2 plume images reveal the sources of the leaks to be two separate locations along the Gryazovets-Ring of the Moscow Region Gas Pipeline. The western plume is coming from a small piece of infrastructure next to the pipeline, and the eastern plume comes from the Gazprom Pereslavskoye compressor station. The blob-like shapes of the Sentinel-2 and Sentinel-3 plumes agree with low wind speeds, which also explains the strong methane enhancement of 500 ppb in Sentinel-5p (7 km × 5.5 km pixel) observations, ~10 ppm enhancement in the 500 m Sentinel-3 observations, and ~50 ppm enhancement in the 20 m Sentinel-2 observations. Sentinel-5p shows enhancement between the two leak sites, but Sentinel-2 and Sentinel-3 do not show the same. The pixel noise for methane detection in Sentinel-2 and Sentinel-3 is significantly higher compared to Sentinel-5p. Therefore, when applying a scene-specific 95-percentile threshold mask, only the pixels showing substantial enhancements, particularly close to the source, are visible in the Sentinel-2 and Sentinel-3 images.

The Sentinel-2 plume detection in the Figure 3C shows a dual-plume structure, which is caused by orthogonal alignment of the sun-to-satellite plane with respect to the wind direction. Dual-plume structures have been reported in high resolution methane plume retrievals by Borchardt et al. (2021) for AVIRIS-NG and Sánchez-García et al. (2021) for WorldView-3. The orthogonal wind alignment enhances the parallax effect introduced by the plume at a certain height and is translated into two different plume projections (see SI Figure 2). Generally, the plume disperses quickly within the atmospheric boundary layer. We can approximate the plume height as half the height of the boundary layer, which can range from several hundred meters to 1-2 kilometers. The plume shown in Figure 3D has a similar observation geometry to the plume in Figure 3C. A weaker dual-plume effect is visible in Figure 3D, possibly due to a small wind direction difference with the location of Figure 3C.

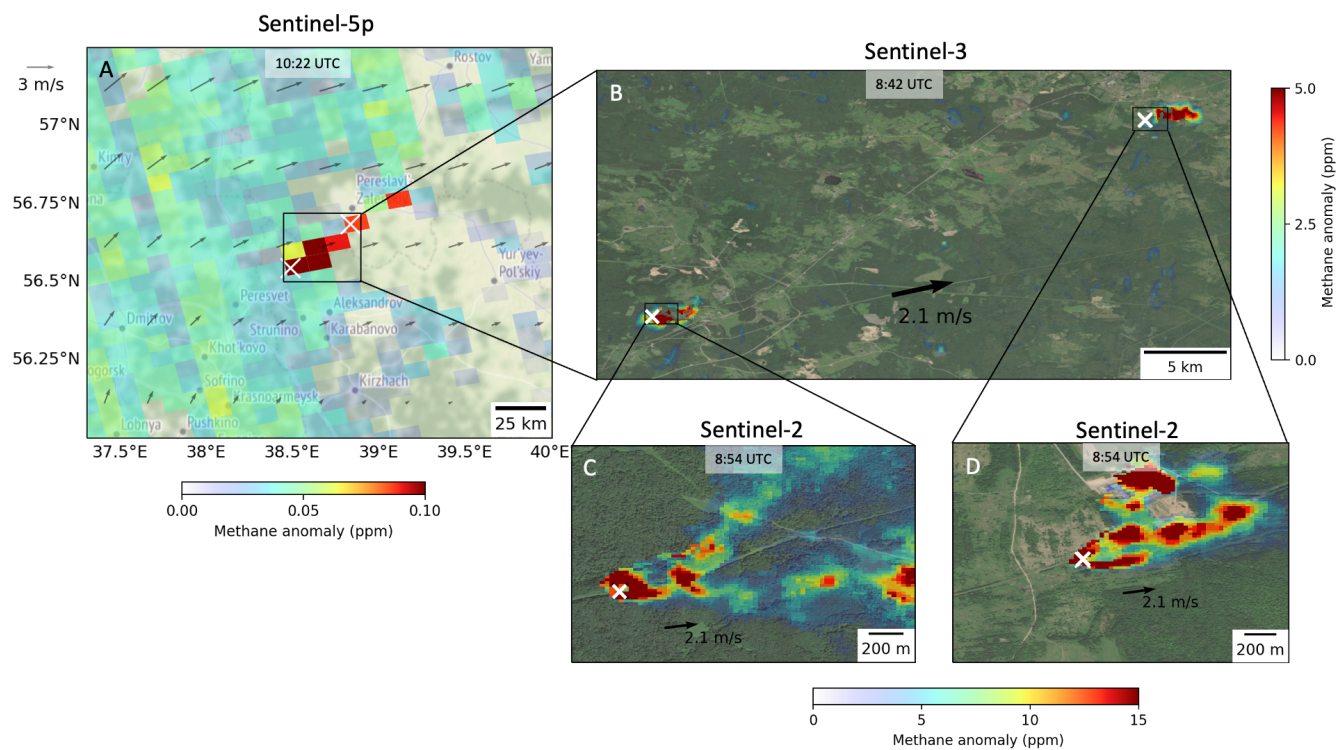

**Figure 3.** Tiered satellite observation of two concurrent methane leaks from a gas pipeline in Russia on June 18, 2021. Sentinel-3 (B) and Sentinel-2 (C & D) observations reveal that the single Sentinel-5p (A) plume is the result of two leaks occurring 30 km apart. The locations of the leaks are marked with crosses in Panel C (56.5446° N, 38.4875° E) and Panel D (56.6850° N, 38.8407° E). The wind vectors show ERA5 10 m winds. The black boxes show sequential zoom-ins by marking the area of Panel B in Panel A and the areas of Panels C and D in Panel B.

The Sentinel-2 and Sentinel-3 plumes are overlaid on ESRI surface imagery, and the Sentinel-5p data is overlaid on OpenStreetMap tiles.

Sentinel-3 has an earlier overpass at the location at 8:01 UTC. On that overpass, no methane plumes are detected at these sites, suggesting that the leaks began, possibly simultaneously, sometime between 8:01 UTC and 08:42 UTC. The next day Sentinel-5p and Sentinel-3 overpasses do not detect a plume at these sites, indicating that the leaks are likely transient events to relieve excess pressure from the gas pipeline and lasted only a few hours.

We quantify source rates for the Sentinel-2 plumes to be 269 ± 67 t/h for the Figure 3C plume and 195 ± 54 t/h for the Figure 3D plume. The source rate quantifications for the Sentinel-3 plumes, 188 ± 70 t/h and 177 ± 65 t/h, respectively, are in agreement with the Sentinel-2 quantification. For the Sentinel-5p plume, the CSF emission estimate is highly uncertain: we find a very large source rate range (10 t/h to 261 t/h) for individual CSF transects because of large variation in methane enhancements along the plume direction. The plumes appear to not be fully developed. They are only two hours 'old' at the Sentinel-5p observation time, which is insufficient time for the emitted gas from the western source to reach the eastern source. This is supported by the fact that the CSF transects near the two sources have high emission estimates (>100 t/h), while the transects further downwind of the sources, including the transects upwind of the eastern source, have lower emission estimates. It is also possible that the large methane enhancements in part of the Sentinel-5p pixel near the source, visible in the Sentinel-2 and Sentinel-3 images, can cause a partial-pixel absorption saturation effect resulting in an underestimate of methane concentration enhancement and emissions estimates (Pandey et al., 2019).

**3.2 Algeria leak**

Figure 4 shows a methane plume detected by Sentinel-5p over Algeria on January 4, 2020 at 12:51 UTC. Originating from the Hassi Messaoud oil/gas field, the plume extends more than 200 km northeast. The orientation of the plume is consistent with ERA5 winds. To identify the precise origin of the Sentinel-5p plume, we sequentially zoom-in on the plume's origin using Sentinel-3 and Sentinel-2. Sentinel-3 observes the scene three hours before

Sentinel-5p (Figure 4B) and locates the source facility, which contains multiple possible plume sources like flare pits and oil/gas wells. Sentinel-2 observes this facility at 10:22 UTC and detects a methane plume from a single oil/gas well, revealing the source of the Sentinel-5p plume.

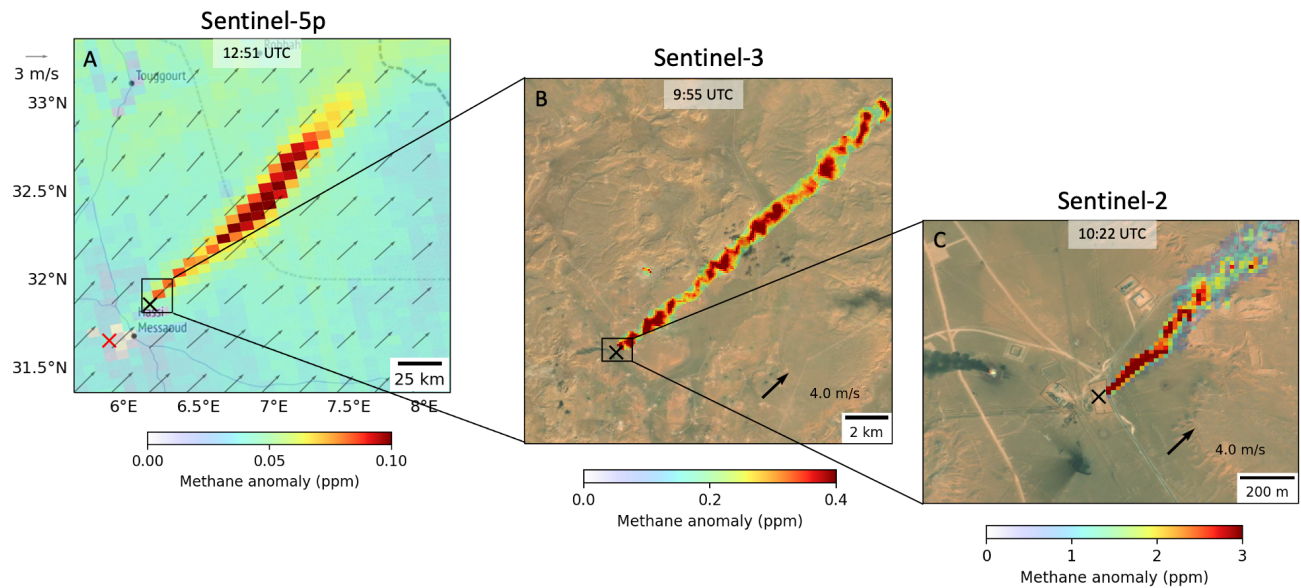

**Figure 4.** Tiered satellite observation of a methane leak from an oil/gas well (31.8639° N, 6.1731° E) in Algeria on January 4, 2020. Panels A, B and C show plume observations by Sentinel-5p, Sentinel-3 and Sentinel-2, respectively. The black 'x' marks the location of the leaking well site. The red 'x' denotes the location (31.6585° N, 5.9053° E) of a concurrently leaking oil/gas site studied by Varon et al. (2021). The black boxes show sequential zoom-ins by marking the area of Panel B in Panel A and the area of Panel C in Panel B. The Sentinel-2 and Sentinel-3 plumes are overlaid on ESRI surface imagery, and the Sentinel-5p data is overlaid on OpenStreetMap tiles.

We investigate the plume activity at the leak site on days before and after January 4. Figure 5 shows clear methane plume detections by Sentinel-5p during January 3–8, with the strongest methane enhancements of > 90 ppb above the respective median backgrounds on each overpass. Sentienl-5p does not observe the scene on January 6 because of clouds. On January 7, Sentinel-5p covers the leak site at the edge of its swath, and the wind is oriented towards the edge of the swath. A few enhanced Sentinel-5p pixels downwind of the source indicate that the leak is active, as confirmed by Sentinel-2 and Sentinel-3 observations. During January 9-19, Sentinel-5p does not observe the site because of clouds. On January 20 and onwards, there is no plume around the site in Sentinel-5p observations.

Sentinel-3 observes plumes on all the Sentinel-5p plume days and on January 7. Sentinel-5p does not detect a plume on January 2 and 20 at the leak site. The lack of plumes on these days is supported by Sentinel-2 observations (SI Figure 3). On the January 9, Sentinel-2 and Sentinel-3 detect large burning activity at the leak site, indicated by very high SWIR band radiances of the two satellites and RGB image of Sentinel-2 (SI Figure 3). The burning at the site continues until February 13, and then the fire spot shifts by 150 m to the north on the well pad. Such activity is typical for gas well blowouts (Maasakkers et al., 2022a), where operators shift the fire site to access the well and control the blowout. Sentinel-3 observes the fire spot at the site until May 14, indicating that the blowout event lasted for about 132 days, with six days of direct gas leakage to the atmosphere and gas burning during the rest of the period.

We quantify source rates for the Sentinel-5p plumes using the CSF method: January 3: 68 ± 28 t/h; January 4: 50 ± 21 t/h. Emissions quantifications for January 5, 7 and 8 are not feasible because of curling plumes due to rotating wind or insufficient coverage. For January 4, we quantify the source rate of the Sentinel-2 plume at 21 ± 6 t/h, and Sentinel-3 plume at 30 ± 12 t/h. The lower Sentinel-2 and Sentinel-3 source rates indicate that the Sentinel-5p plume likely represents an accumulation of methane emitted from additional nearby point and/or diffuse sources from Hassi Messaoud. A mass-balance method like CSF cannot differentiate between the emissions from these sources and the main leak site due to the coarse spatial resolution of the Sentinel-5p data. High-resolution satellite data have identified multiple methane-leaking sites in Hassi Messaoud, but they do not cause methane plumes as large as the one observed during January 3–8 (Ehret et al., 2022). Varon et al. (2021) showed that another leak site in Hassi Messaoud (the red cross in Figure 4A) was active between October 9, 2019 and August 9, 2020 and emitted at an average source rate of 9.3 t/h. SI Figure 4 shows Sentinel-2 plume detection for the Varon et al. (2021) site on January 4 (source rate estimate = 6 ± 3 t/h). The source rate agreement of Sentinel-5p with Sentinel-2 and Sentinel-3 improves after subtracting the Varon et al. (2021) site's contribution from the Sentinel-5p source rate.

Overall, these case studies show that each satellite within the tiered observing system provides unique information on leaks. The daily global coverage of Sentinel-3 informs about the duration of the leaks and its high spatial resolution helps identify the sources of the Sentinel-5p plumes. Sentinel-2 lacks daily global coverage but gives a more precise source location with its 20 m spatial resolution data. The satellites also have complementary strengths for source rate quantification. Sentinel-5p gives the total emissions from multiple nearby point and diffused sources.

Sentinel-2 and Sentinel-3 cannot observe the diffuse emissions, but they can quantify individual sources rate from important super-emitter point sources.

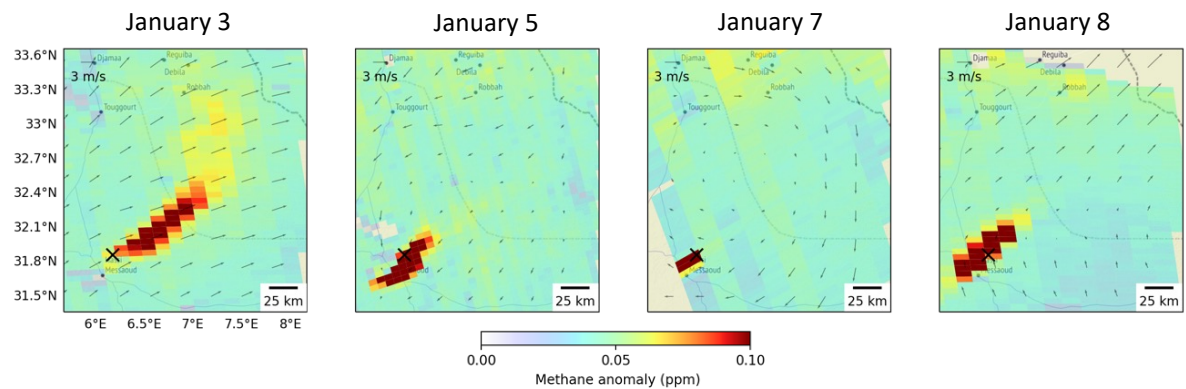

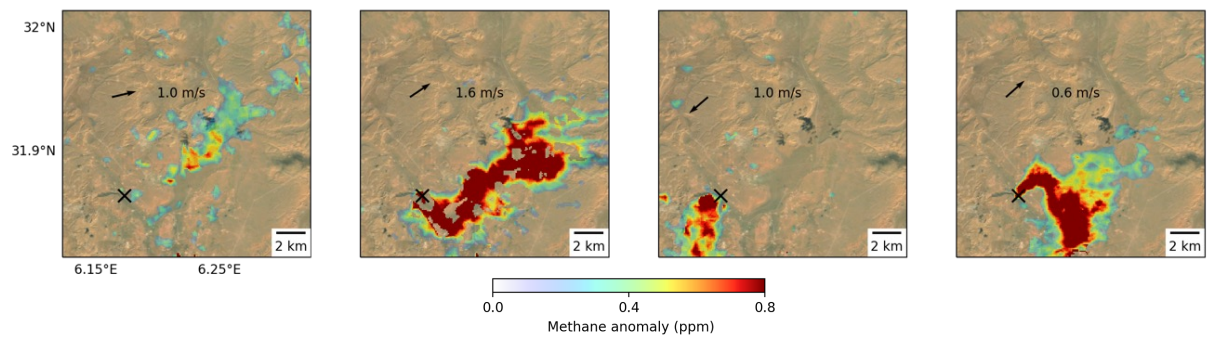

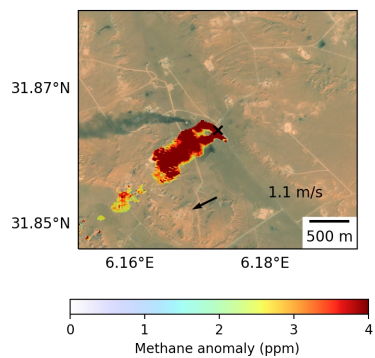

**Figure 5.** Observations of methane plumes from the Algerian point source during January 3–8, 2020. The top, middle, and bottom rows show observations by Sentinel-5p, Sentinel-3 and Sentinel-2, respectively. Detections on January 4 are shown in Figure 4. On January 6, the leak was not observed by the daily-overpass Sentinel-5p and Sentinel-3 satellites because of clouds. The Sentinel-2 and Sentinel-3 plumes are overlaid on ESRI surface imagery, and the Sentinel-5p data is overlaid on OpenStreetMap tiles. The spatial domains of the respective satellite images are similar to those shown in Figure 4.

## 4 Sentinel-3 detection limit

The SWIR-2 band of Sentinel-3 (~50 nm) is narrower than that of Sentinel-2 (~200 nm) and lies at a stronger methane absorption wavelength (Figure 1.A). Therefore, SWIR-2 sensitivity of Sentinel-3 is better than that of Sentinel-2 by up to a factor of two (Figure 1.B), while the SWIR-1 sensitivity of the two satellites is similar. The fractional methane absorption signal of the MBMP method is the ratio between the SWIR-1 and SWIR-2 signals, and it is also better for Sentinel-3 by a factor of 2. However, this does not translate into a more effective methane leak detection by Sentinel-3 because the enhancements in a plume are much stronger when sampled at 20 m Sentinel-2 pixels in comparison to 500 m Sentinel-3 radiance observation pixels. For example, the average of the Sentinel-3 plume enhancement (0.4 ppm) is 12 times lower than the Sentinel-2 plume enhancement (4.9 ppm) for the Korpezhe plume shown in Figure 2. We compared average plume enhancements rather than peak pixel enhancements here because multi-band satellites use the plume-like shape of the observed enhancement for plume detection. The peak single-pixel enhancement is not useful for detection because of large noise in methane retrievals of multi-band satellites. The enhancement difference between the Sentinel-2 and Sentinel-3 Korpezhe plumes roughly translates to a 6 times worse Sentinel-3 detection limit after accounting for the superior sensitivity of Sentinel-3's SWIR-2 band.

The ratio of plume enhancement between Sentinel-3 and Sentinel-2 depends on wind conditions and the level of noise within the spatial domain of the scene, as the extent of the plume is typically delineated by a noise threshold (for example, a 95-percentile filter). We find that the methane retrieval noise in Sentinel-2's 20 m pixels is typically 3-4 times higher than that in Sentinel-3's pixels of 100 m size (result of regridding of the 500 m SWIR pixels of Sentinel-3; see Section 2.2). This leads to visibly shorter methane plumes for Sentinel-2, as well as 95-percentile plume masks. The noise within a scene is influenced by factors such as the signal-to-noise ratio (SNR) of different SWIR bands, observation geometry, solar zenith angle, and the presence of stripes and noise related to surface features. Additionally, the detection limit is impacted by the area required for plume identification. Given its smaller spatial domain size requirement, Sentinel-2 offers an increased likelihood of a scene free of noise, water bodies, and other surface-related artifacts.

We conduct an empirical test of Sentinel-3's source rate detection limit by analyzing three test sites known for persistent Sentinel-2 plume observations: (1) Korpezhe Compression station in Turkmenistan; (2) Algerian leak site; and (3) Parks compressor station in the Permian basin, USA. Sites (1) and (2) were investigated by Varon et al. (2021), while site 3 was investigated by Varon et al. (2022). These studies quantified the source rates from Sentinel-2 plume observations using the Integrated Mass Enhancement (IME) method with local GEOS-FP 10 m wind speed data (Molod et al., 2012). Using these Sentinel-2 source rates, we discern a range of minimum plume detections by Sentinel-3 at these diverse locations. The outcomes of our validation for the three sites are shown in Table 1 and Figure 6.

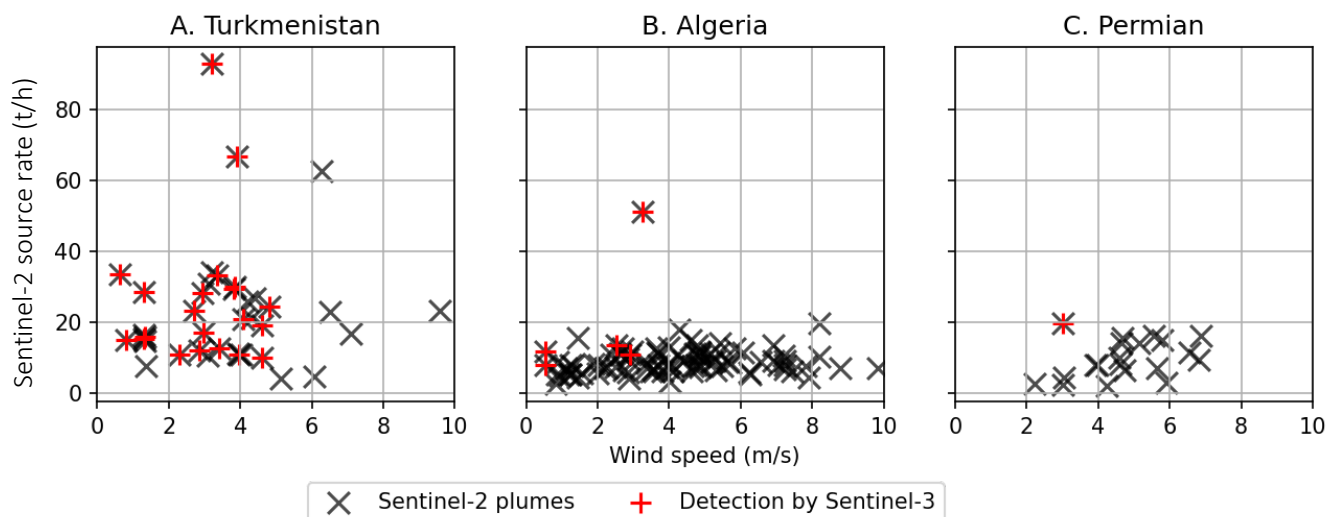

**Figure 6.** Detection limit analysis for Sentinel-3 at the three test sites. These panels display the Sentinel-2-based plume source rates on the y-axis and the 10 m wind speeds from GEOS-FP on the x-axis. Sentinel-2 plumes (black 'x') from days with successful Sentinel-3 methane retrievals are shown. The co-detections of plumes by Sentinel-3 are indicated by red "+".

When examining each test site, we first identify successful methane retrievals by Sentinel-3 on days when a plume was observed by Sentinel-2. The absence of Sentinel-3 methane retrievals on a particular day could be due to a variety of factors, including the Sentinel-2 detections that occurred before the launch of both Sentinel-3A (February 2016) and 3-B (April 2018, both instruments are needed for daily global coverage), cloud cover during the Sentinel-3 overpass time, instrument issues, and the presence of a large flare near the test site. We then visually inspect each methane retrieval image by Sentinel-3 for the presence of methane plumes.

**Table 1.** Investigation of Sentinel-3 detection limits relative to Sentinel-2 plume detections.

| Test Sites | Sentinel-2 Plume Detection Period | Successful Sentinel-3 Methane Retrieval Days | Number of Sentinel-3 Plume Detections | Sentinel-2 Source Rate* average [range] (t/h) | | Average GEOS-FP 10m Wind Speed (m/s) | |
|---|---|---|---|---|---|---|---|
| | | | | Sentinel-3 non-detection | Sentinel-3 detection | Sentinel-3 non-detection | Sentinel-3 detection |
| Algeria (31.6585° N, 5.9053° E) | October 2019- August 2020 | 88 | 5 | 9 [3-19] | 19 [8-51] | 4.3 | 1.9 |
| Permian (31.731° N, 102.042° W) | April 2018- October 2020 | 20 | 1 | 9 [2-16] | 20 | 4.9 | 3.0 |
| Turkmenistan (38.4939° N, 54.1977° E) | May 2016- November 2020 | 39 | 24 | 19 [4-62] | 26 [10-93] | 4.4 | 2.9 |

At three test sites, methane plumes were detected by Sentinel-2 with different frequency and intensity, and Sentinel-3 detects fraction of the Sentinel-2 plumes. The Algerian site had five detections by Sentinel-3 out of 88 successful retrievals, Permian only had one detection, while Turkmenistan had 24 detections out of 39 successful Sentinel-3 retrievals. In the Permian Basin, the one detected plume measured approximately 20 t/h. However, at the Algerian and Turkmenistan test sites, plumes as small as 10 t/h were detected. The weakest winds in the Permian Basin are only 3 m/s, compared to the other two sites where wind speeds are as low as 1 m/s. The average wind speed in the Permian Basin is also higher: 4.8 m/s in Permian vs 4.2 m/s in Algeria and 3.7 m/s in Turkmenistan.

The largest plume that Sentinel-3 failed to detect at the Turkmenistan site (on 20 May, 2018) has a source rate of 62 t/h. This may be due to the leak beginning very close to the overpass times of Sentinel-3 (6:29 UTC) and Sentinel-2 (7:10 UTC), resulting in the methane plume not extending over a sufficient area to create observable plumes within Sentinel-3's 500 m pixels. Furthermore, when methane enhancement covers only a small portion of the Sentinel-3 observation pixel, it can result in an underestimation of the methane absorption signal due to the partial pixel enhancement effect (Pandey et al., 2019).

Our investigation indicates that Sentinel-3's detection limit ranges between approximately 8 and 20 t/h, subject to differences across various scenes. The Sentinel-3 detection limit is strongly influenced by the wind speed. At high wind speeds, even larger source rate plumes may not be detected because of low methane column enhancement due to the stronger dispersion of the plume. This is in addition to the impact of surface albedo, solar zenith angle, and surface feature-related noise.

Gorroño et al. (2022) investigated the detection limit of Sentinel-2 using end-to-end radiative simulations. They illustrated that the Sentinel-2 detection limit can vary by up to a factor of 5, depending on the observational conditions. Ehret et al. (2022) empirically confirmed the variation in Sentinel-2 detection limits for different scenes through a global assessment of Sentinel-2 plume detections. In line with these findings for Sentinel-2, we observe that Sentinel-3's detection limit deteriorates depending on the observational conditions of a location. The better detection limit for Sentinel-2 at the Permian site (2 t/h), compared to the Turkmen (4 t/h) and Algerian sites (3 t/h), is likely due to improvements and alterations in the MBMP method applied between Varon et al. (2021) and Varon et al. (2022). In the former, fixed reference days were used, whereas the latter study chose reference days for each image using correlation with the main day's non-methane bands, similar to our approach in this study.

## 5 Sentinel-3 within a tiered observation approach

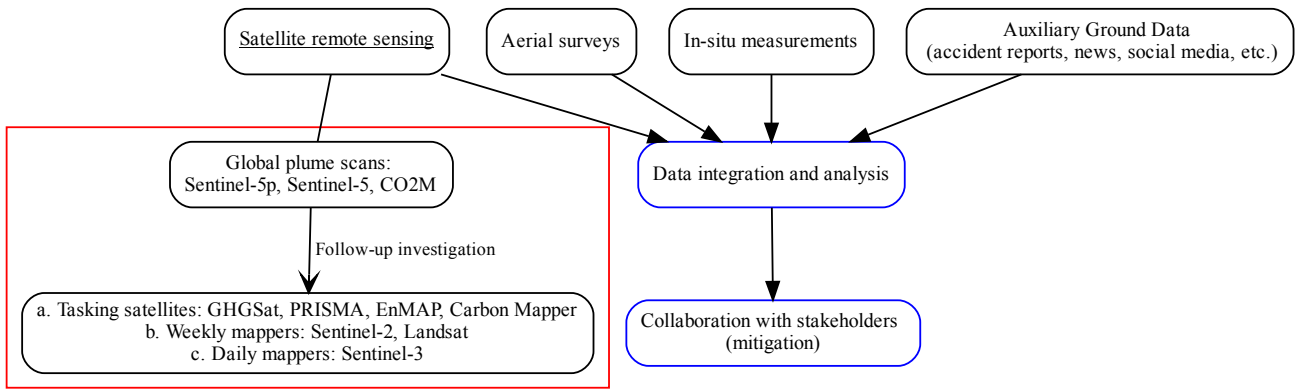

**Figure 7.** Illustration of a tiered approach for detecting methane leaks. The satellite remote sensing component of the approach is highlighted by the red box, while the blue rounded rectangles emphasize the processes following detection, culminating in mitigation. Further information about the satellites mentioned in this flowchart can be found in Jacob et al. (2022). The examples of various types of instruments provided do not constitute exhaustive lists.

Here we discuss how Sentinel-3's detection capability for large methane plumes, with daily global coverage and high resolution, contributed to a tiered methane leak observation approach. A tiered methane leak observation approach (Figure 7) combines different satellite and surface-based instruments, along with auxiliary sources of information, to allow for more comprehensive monitoring of methane leaks. Recent studies have underscored the efficacy of this approach for quantifying large accidents (well blowouts) in the U.S., utilizing information from different (targeted) satellite platforms as well as supporting information from reports, news, satellite-observed flaring, and ground-based local air quality observations (Pandey et al. 2019; Cusworth et al., 2021; Maasakkers et al. 2022a). Especially when surface-based observations and auxiliary information are unavailable, such as in remote areas, the satellite remote sensing component of the tiered observation approach becomes essential.

Observations from Sentinel-5p are now routinely scanned to detect leaks globally (Lauvaux et al., 2022; Schuit et al., 2023). This approach will also be implemented for upcoming high-precision global mappers, including Sentinel-5 and CO2M satellites. Using Sentinel-5p, origins of detected plumes can only be pinpointed within a few kilometers (Maasakkers et al. 2022b). Moreover, it is becoming increasingly clear that plumes detected by Sentinel-5p often encompass emissions from multiple leaks (Schuit et al., 2023). As a result, satellites with higher spatial resolution are required to locate the sources of methane leaks. This can be achieved through a tip-and-cue strategy, where satellites with higher resolution such as GHGSat, PRISMA, EnMAP, and Carbon Mapper are tasked to survey the area approximated by Sentinel-5p as the plume's origin. However, many leaks of short duration do not allow enough time to fully exploit the capabilities of tasking satellites. For instance, the two large pipeline leaks in Russia analyzed in this study likely lasted for just a few hours (Section 3.1). In such scenarios, multi-band global mapping instruments like Sentinel-2 and Sentinel-3 can be beneficial with their ability to provide global coverage in less than a week or even a day, and detect sources as small as 1 t/h to 10 t/h. A unique advantage of using Sentinel-3 in this tiered approach is its capacity to deliver daily global coverage at high spatial resolution. As a result, it can detect substantial methane sources that fall outside of the coverage of other high-resolution satellites and monitor sources on a daily basis.

## 6 Summary

We have demonstrated successful detection of methane plumes from Sentinel-3 SWIR band data, which provide daily global coverage with a spatial resolution of 500 m. We retrieved methane enhancements from Sentinel-3 and Sentinel-2 SWIR data using the multi-band multi-pass method. By comparing with the methane plume detections of Sentinel-2, we reported Sentinel-3 plume detections of source rates down to 8-20 t/h, depending on location and wind conditions. Using two case studies, we demonstrated the methane leak identification and monitoring capabilities of Sentinel-3 as well as the complementarity of using Sentinel-5p, Sentinel-3 and Sentinel-2 together to characterize methane super-emitter leaks. For a Sentinel-5p plume detection in Russia, Sentinel-3 showed that two leaks occurred simultaneously 30 km apart along a pipeline but resulted in a single plume in Sentinel-5p data because the leak sites were located along the wind direction. In Hassi Messaoud, Algeria, we found that an oil/gas facility emitted massive amounts of methane during a 6-day period followed by four months of burning of the leaking gas. Sentinel-2 identified the precise location of the leak to be an oil/gas well. Sentinel-3 showed the leak continued for 6 days, resulting in massive enhancements in Sentinel-5p data.

We demonstrated the advantages of the tiered observation approach of combining information from satellites with different coverages and spatial resolutions to monitor super-emitter leaks. Within the tiered observation approach, the daily global coverage of Sentinel-3 informs about the duration of the leaks and its high spatial resolution helps to identify the sources of the Sentinel-5p plumes in absence of coverage from higher spatial resolution satellites.

*Data availability*: The Sentinel-5p methane dataset of this study is available for download at ftp://ftp.sron.nl/open-access-data-2/TROPOMI/tropomi/ch4/ (version 19_446, accessed 10-08-2022). Sentinel-2 and Sentinel-3 data are available at Copernicus Open Access Hub (https://scihub.copernicus.eu/dhus/; accessed 27-10-2022). ERA5 wind data is available from Copernicus Climate Data Store (https://cds.climate.copernicus.eu/cdsapp; accessed 27-10-2022).

*Acknowledgement:* Part of the research was carried out at the Jet Propulsion Laboratory, California Institute of Technology, under a contract with the National Aeronautics and Space Administration (80NM0018D0004). S.P. acknowledges funding through the GALES (GAs Leaks from Space) project (Grant 15597) by the Dutch Technology Foundation, which is part of the Netherlands Organisation for Scientific Research (NWO). P.T. is funded by the TROPOMI national program from the NSO.

# Supplementary Information (SI) for "Daily detection and quantification of methane leaks using Sentinel-3: a tiered satellite observation approach with Sentinel-2 and Sentinel-5p"

Sudhanshu Pandey[1,2], Maarten van Nistelrooij[1], Joannes D. Maasakkers[1], Pratik Sutar[1], Sander Houweling[3], Daniel J. Varon[4], Paul Tol[1], David Gains[5], John Worden[2], Ilse Aben[1,3]

[1] SRON Netherlands Institute for Space Research, Leiden, the Netherlands
[2] Jet Propulsion Laboratory, California Institute of Technology, Pasadena, CA, USA
[3] Department of Earth Sciences, Vrije Universiteit Amsterdam, Amsterdam, the Netherlands
[4] School of Engineering and Applied Sciences, Harvard University, Cambridge, United States
[5] GHGSat, Inc., Montréal, H2W 1Y5, Canada

*Correspondence to*: Sudhanshu Pandey (sudhanshu.pandey@jpl.nasa.gov)

## 1 Satellite data

### 1.1 Sentinel-2

Sentinel-2 is a European Earth observation mission designed to provide operational data products for environmental risk management, land cover classification, land change detection, and terrestrial mapping (Drusch et al., 2012). It comprises two satellites positioned 180° out of phase in the same sun-synchronous orbit, with an Equator-crossing time of 10:30 (local solar time) at the descending node. Sentinel-2A was launched in June 2015, and Sentinel-2B in March 2017. Each satellite carries a multi-band instrument that continuously sweeps the Earth's surface in 13 spectral bands from the visible to the shortwave infrared (SWIR) at 10–60 m pixel resolution over a 290 km cross-track swath. The twin satellite configuration enables full global coverage every five days. We use Sentinel-2 level 1C (L1C) top-of-atmosphere reflectance data of SWIR-1 (~1560–1660 nm) and SWIR-2 (~2090–2290 nm) bands to retrieve methane column enhancements.

### 1.2 Sentinel-5p

TROPOMI is the single instrument onboard ESA's Sentinel-5p satellite. It was launched in 2017 in a sun-synchronous orbit at 824 km altitude with an equator-crossing time of 13:30 (local solar time) at the ascending node (Veefkind et al., 2012). It is a push-broom imaging spectrometer recording spectra along a 2600 km swath while orbiting Earth every 100 min, resulting in daily global coverage. Total column methane is retrieved with nearly uniform sensitivity in the troposphere using earthshine radiance measurements from its shortwave infrared channel of 2305–2385 nm spectral range and 0.25 nm spectral resolution. These measurements have a ground pixel size of 7 × 5.5 km² at the nadir, with larger pixels towards the edges of its swath. We use version 19 of the SRON scientific methane retrieval product (Lorente et al., 2022). We correct the along-track stripes in the data, resulting from varying calibration uncertainty across the track, following Borsdorff et al. (2018).

**2 Methane retrieval method for Sentinel-3 and Sentinel-2**

We use the multi-band multi-pass (MBMP) method to retrieve methane column enhancements ($\Delta\Omega$) from the SWIR band measurements (Varon et al., 2021). We first calculate the fractional change in TOA radiances ($\Delta R$) using a combination of the SWIR radiances ($R_{SWIR1}$, $R_{SWIR2}$):

$$\Delta R = \frac{c\, R_{SWIR2}}{R_{SWIR1}} - 1,$$

where $c$ is a scaling factor that adjusts for scene-wide changes in brightness between the two bands. It is calculated by least-squares fitting of all $R_{SWIR2}$ values against all $R_{SWIR1}$ values using a first-order linear regression with zero intercept. The systematic errors in $\Delta R$ due to wavelength separation between the bands are corrected by subtracting $\Delta R$ over the same scene from a reference image.

To convert $\Delta R$ to $\Delta\Omega$, we calculate fractional absorption ($m$) using a radiative transfer model:

$$m(\Delta\Omega) = \frac{T_{SWIR2}(\Delta\Omega)}{T_{SWIR2}(0)} - \frac{T_{SWIR1}(\Delta\Omega)}{T_{SWIR1}(0)}.$$

$T$ is the simulated TOA radiance integrated over the spectral range of SWIR bands of Sentinel-3/2. We use a 25-layer, clear-sky radiative transfer model to calculate $T$. The model accounts for molecular absorption but not thermal emission or scattering. We simulate radiances at 0.025 nm spectral resolution and integrate them over the SWIR band spectral windows. The model relies on U.S. Standard Atmosphere vertical profiles of pressure, temperature, air density, water vapor, $CO_2$, and background methane (Anderson et al., 1986). Absorption line strengths for the trace gases are obtained from the HITRAN 2016 database (Gordon et al., 2017). The radiative transfer model accounts for solar zenith angle ($sza$) and instrument zenith angle ($iza$). Methane enhancement is presumed to be in the lowest layer of the model of 1 km thickness. The radiative transfer model accounts for absorption by CO, $N_2O$, $CO_2$, and $H_2O$; and the air mass factor ($amf$). $amf$ depends on the observation geometry as $amf = \frac{1}{\cos(sza)} + \frac{1}{\cos(iza)}$. We determine $\Delta\Omega$ by solving for $\Delta R = m(\Delta\Omega)$. To save on computation, we calculate lookup tables for $m$ for a range of $\Delta\Omega$ and $amf$ values. Information in the coarse spectra measured by Sentinel-3 (and Sentinel-2) is low, so successful methane retrievals require strong methane enhancements and small gradients in the concentration fields of non-methane gases that influence the SWIR bands (like $H_2O$ and $CO_2$). Such favorable conditions often occur with large methane leaks at oil/gas sites, unlit or malfunctioning flare installations, or compressor stations venting natural gas.

# 3 Cross-sectional flux (CSF) method

The CSF method relates source rate $Q$ (g/s) and effective wind speed $U_{eff}$ (m/s) as follows:

$$Q = C \times U_{eff},$$
$$C = \frac{1}{n} \sum_{j=1}^{n} \Delta\Omega\left(x_j, y\right) dy,$$

where $C$ (g/m) is line-integrated methane column enhancement computed by sampling the plume pixels along transects orthogonal to the plume direction (y-axis) in the downwind of the source (x-axis). The sampled pixels are integrated across each transect line. We take the mean of source rates calculated for individual transects ($j = 1 \ldots n$, where $n$ is the number of transects) between the source and the end of the plume. Methane column enhancement for a plume pixel is calculated by subtracting the median of methane columns in the Sentinel-5p observation scene.

For Sentinel-5p plumes, we calculate $U_{eff}$ from ERA5 pressure-weighted average boundary layer wind speed ($U_{bl}$) using the relation from Varon et al. (2019): $U_{eff} = (1.05 \pm 0.17)\ U_{bl}$. This relation was estimated by performing a linear regression between the true source rates and the calculated sources rate from the CSF method application on simulated Sentinel-5p plumes obtained from WRF-CHEM model runs.

For Sentinel-3 and Sentinel-2, $U_{eff}$ is estimated as a function of ERA5 10 m wind speed ($U_{10}$). We calibrate this function using an ensemble of large eddy simulations (LES) of methane point sources following Varon et al. (2021). The LES ensemble includes five simulations over a $9 \times 9 \times 2.4$ km$^3$ domain at 25 m resolution, with different settings for sensible heat flux (100-300 W/m$^2$) and mixed layer depth (500-2000 m). We generate Sentinel-3 pseudo-observations from the LES by degrading the spatial resolution to 100 m, scaling the model source rate to reflect random values in the range 50-100 t/h, and applying Gaussian noise with scene-dependent amplitude to individual plume snapshots. To construct Sentinel-2 pseudo-observations, a similar procedure is applied, but the LES resolution was kept at 25 m. We then construct plume masks for the synthetic observations by selecting pixels with methane column enhancements above the 90$^{th}$-95$^{th}$ percentile for the scene and smoothing the mask with a $3 \times 3$ pixels median filter.

By applying the CSF method on the synthetic plumes with varying noise levels and mask percentile thresholds, we obtain a linear dependence: $U_{eff} = \alpha\ U_{10}$ for both satellites, with $\alpha = 1.11$– $1.19$ for Sentinel-3, and $\alpha = 1.16$– $1.37$ for Sentinel-2, depending on the noise level and mask percentile threshold. The uncertainties of the source rate estimates are calculated as the sum-in-quadrature of (1) measurement error, (2) wind error, and (3) quantification method error. We use $U_{eff}$- $U_{10}$ slope ranges as the CSF method error. Wind error is calculated as the standard deviation of wind speeds across 5 hours near the satellite overpass time. The standard deviation of the emission estimates at different transects is used as measurement error.

# 4 Image denoising

## 4.1 Stripes removal

Stripes are a common issue in remote sensing data. We find oblique stripes in some Sentinel-3 MBMP images, i.e., stripes not aligned with the horizontal or vertical direction of the image data array. SI Figure 6 shows an example of stripes in an MBMP image. To remove the stripes, we first determine the angle of the stripes by applying a 3 × 3 pixels median filter to generate a smoothened image and subtract it from the methane retrieval image. The smoothened image gives an image with enhanced stripes. We calculate this stripe-enhanced image's two-dimensional fast Fourier transform (FFT), followed by shifting the FFT image horizontally to center the FFT image at zero frequency, creating symmetry around its center. The stripe angle is given by the direction orthogonal to the line joining the two maximum values of the image (Liu et al., 2018). We rotate the image clockwise by the stripe angle to align the stripes with the horizontal direction. We then subtract a row-wise median from the rotated image to remove stripes and reorient the image by rotating anticlockwise by the stripe angle. This stripe-removal procedure can be repeatedly applied to remove stripes in multiple directions.

## 4.2 Surface noise masking

We mask noise due to surface features, clouds, and cloud shadows that are visible in radiance images of Sentinel-3's VNIR bands (550, 650, and 870 nm) of the main and reference images using a feature masking method based on structural similarly index (*SSIM*, Wang et al., 2004). *SSIM* estimates the similarity between two images based on luminance, contrast, and structure. For methane retrieval image $a$ and VNIR band image $b$, SSIM is calculated as follows:

$$SSIM\,(a,b) = \frac{(2\mu_a\,\mu_b + c_1\,)(2\sigma_{ab} + c_2)}{(\mu_a^2 + \mu_b^2 + c_1)\,(\sigma^2{}_a + \sigma_b^2 + c_2)}.$$

Here, $\mu$ is the mean, $\sigma$ is the standard deviation, and $\sigma_{ab}$ is the cross-correlation between $a$ and $b$. $c_1$ and $c_2$ are small constants to stabilize the results. SSIM values range between 0 and 1. We scan an MBMP image to calculate local *SSIM* values w.r.t VNIR bands over each set of 3 × 3 pixels. Then, pixel sets with SSIM values larger than a threshold value for any of the VNIR bands are masked out. SI Figure 7 shows an example of noise feature masking using this procedure.

**Figures**

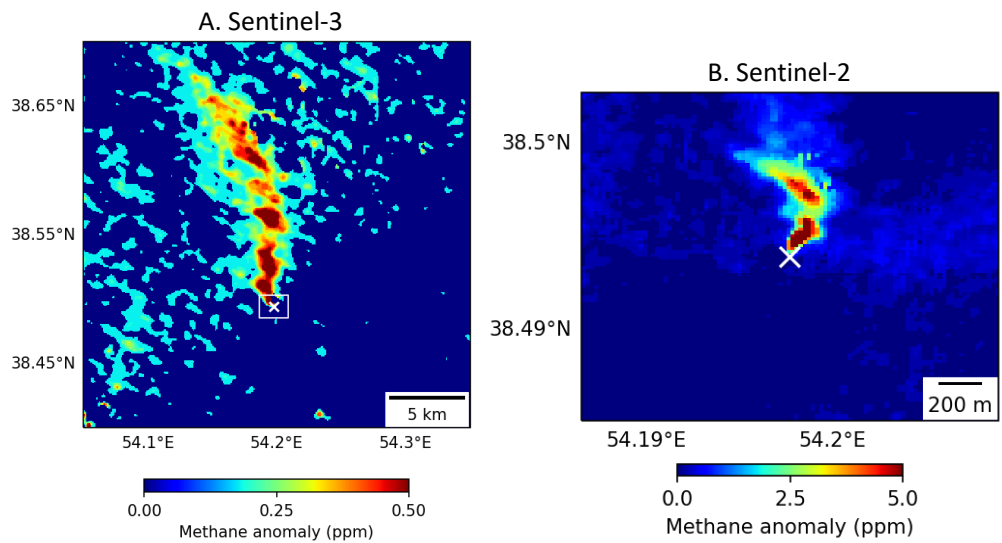

**Figure 1.** Methane concentration enhancement image of Sentinel-3 (A) and Sentinel-2 (B) on 29 June 2018 from a point source near the Korpezhe compressor station (white 'x'; location: 38.4939° N, 54.1977° E) in Turkmenistan.

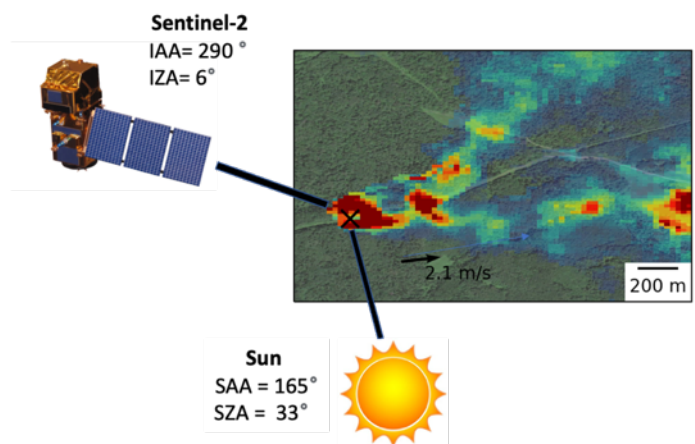

**Figure 2**. Observation geometry **of** the Figure 3C Sentinel-2 plume in the main text. The leak's source location (56.5446° N, 38.4875° E) is marked with a cross. The dual-plume structure results from the orthogonal alignment of the sun-to-satellite plane relative to the wind direction. Values of instrument solar azimuth angle (SAA), solar zenith angle (SZA), instrument azimuth angle (IAA), and instrument zenith angle (IZA) are given. The plumes are overlaid on ESRI surface imagery.

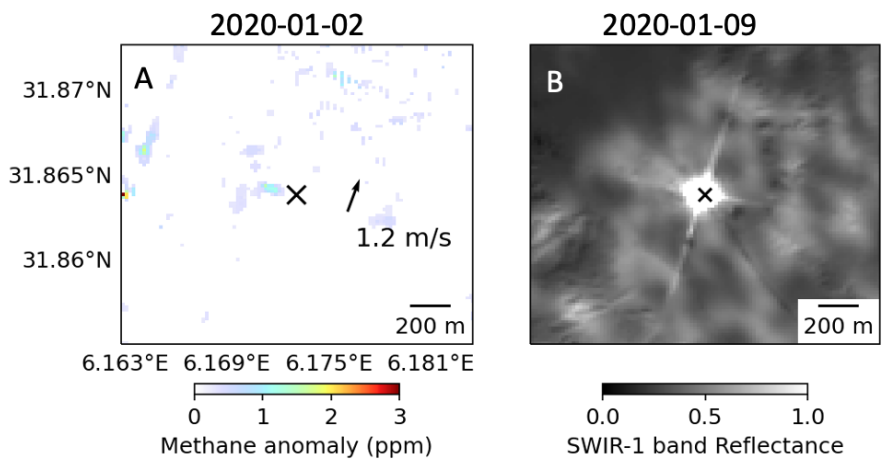

**Figure 3.** (A) Sentinel-2 methane enhancement retrieval on January 2 showing no plume at the Algerian leak site. (B) Sentinel-2 SWIR-1 TOA reflectance on January 9 shows the burning of natural gas as a bright spot at the center of the image.

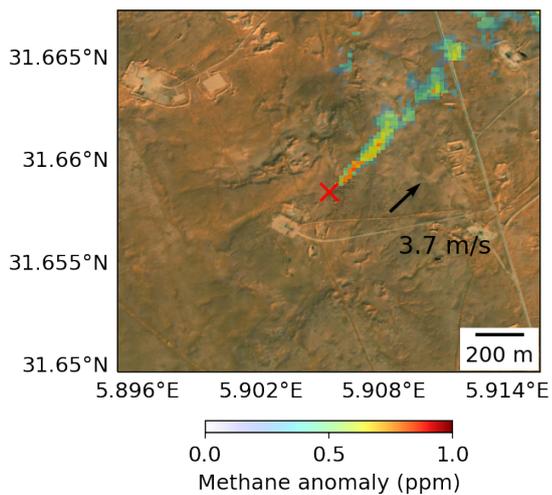

**Figure 4**. Sentinel-2 plume detection for the Varon et al. (2021) Algerian leak site on January 4, 2020. The plume is overlaid on ESRI surface imagery.

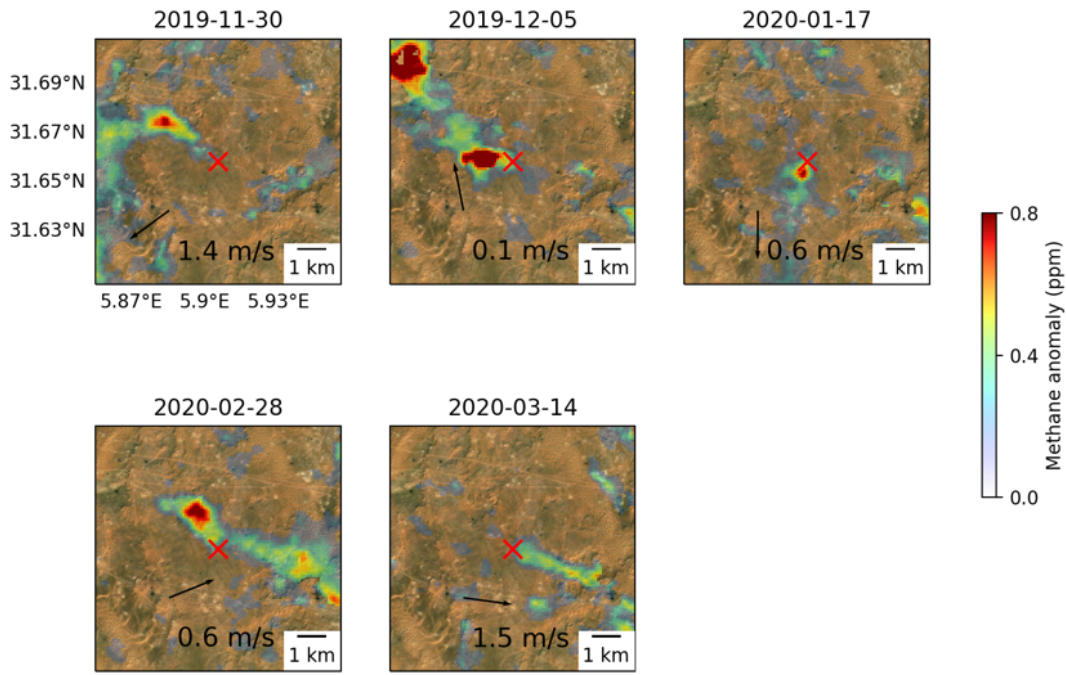

**Figure 5**. Sentinel-3 methane plume retrievals for the Varon et al. (2021) Algerian leak site with ERA5 10 m winds. The retrievals are noisy due to surface albedo variations and flares. The plumes are overlaid on ESRI surface imagery.

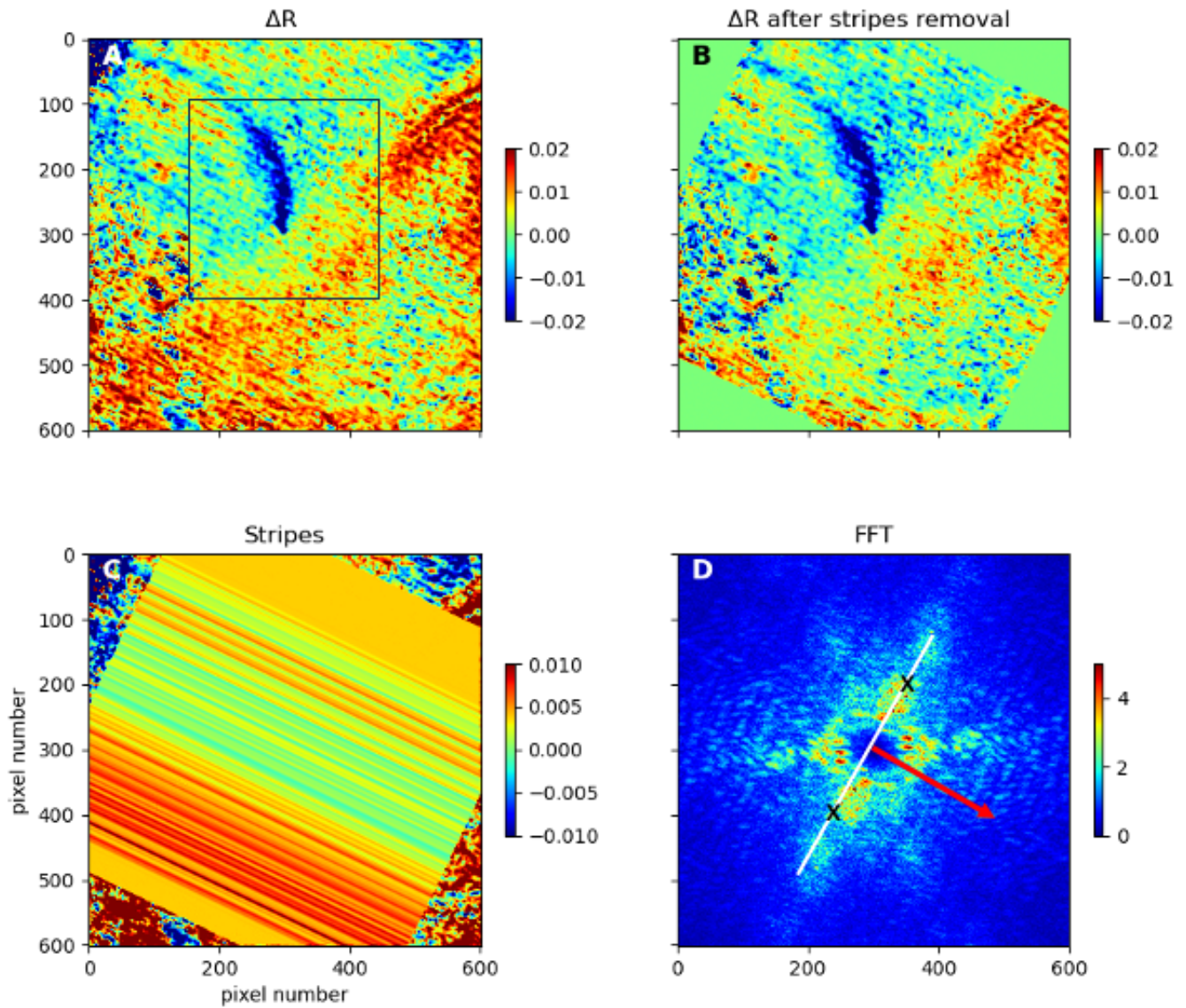

**Figure 6.** Oblique stripe removal for the MBMP ΔR image of the Korpezhe compressor station plume from Figure 2 of the main text. Panels A and B show ΔR before and after stripe removal, respectively, and Panel C shows the difference between A and B. Panel D shows the absolute values of the 2-D fast Fourier transform, shifted horizontally to put zero frequency at the center of the image. The white line connects the largest two FFT values marked by black crosses. The stripe direction (red arrow) is perpendicular to the white line. The black rectangle in panel A delineates the spatial extent of the Sentinel-3 image displayed in Figure 2 and SI Figure 7.

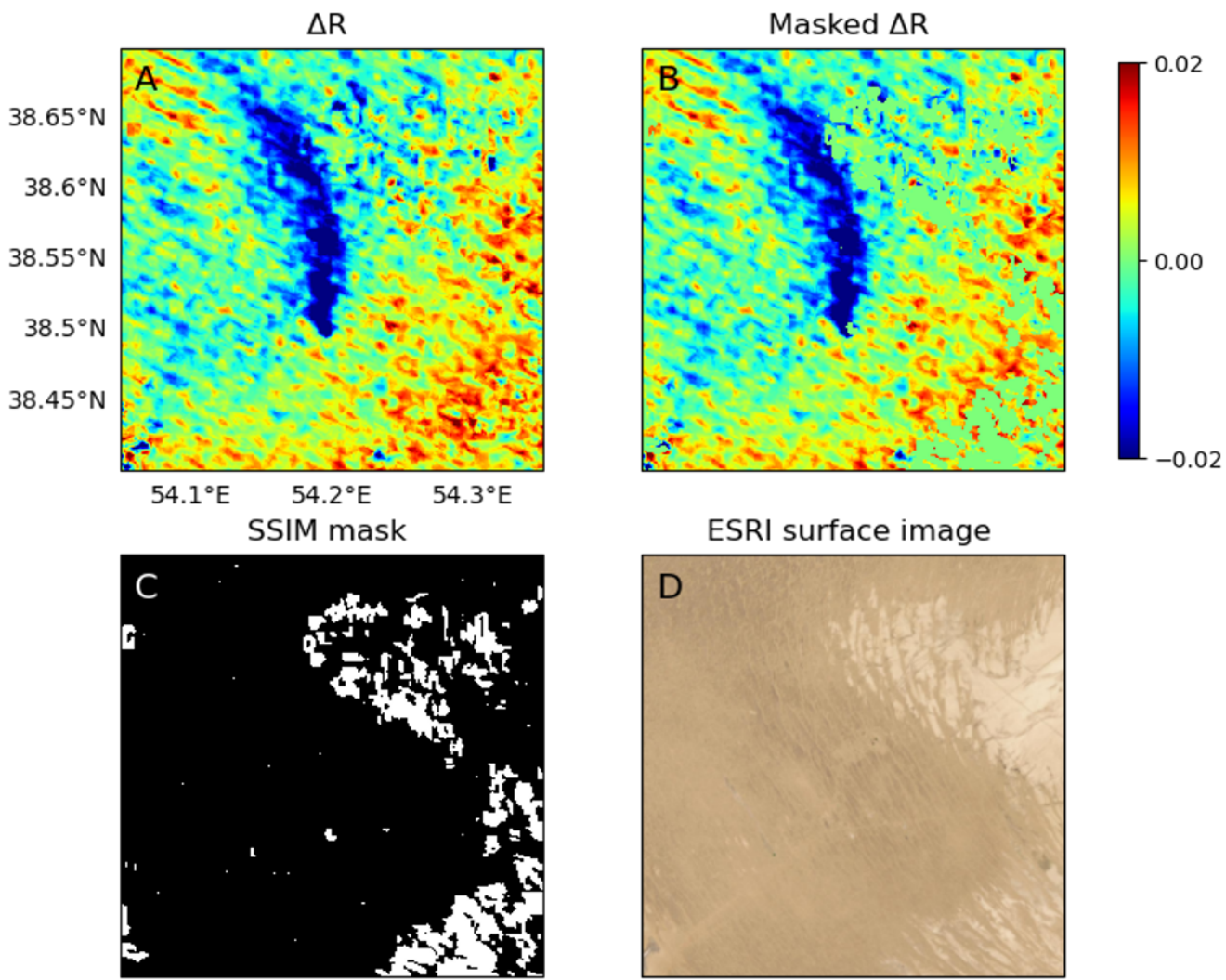

**Figure 7.** SSIM-based denoising of the MBMP Δ$R$ image of the Korpezhe compressor station plume from Figure 2 of the main text. Panel C shows the feature mask with white pixels, which are set to zero in panel B. Panel D displays ESRI surface imagery for the scene. Notice that the image's lower right corner is masked due to noise from surface texture changes.